\documentclass[preprint,superscriptaddress,showpacs,
               preprintnumbers,amssymb,aps,prl]{revtex4}
\usepackage{graphicx,dcolumn,bm,amsmath}
 
\begin{document}
 
\title{Influence of hydrodynamic interactions
on lane formation in oppositely charged driven colloids}

\author{M. Rex}
\affiliation{Institut f\"ur Theoretische Physik II: Weiche Materie,
Heinrich-Heine-Universit\"at D\"{u}sseldorf, 
Universit{\"a}tsstra{\ss}e 1, D-40225 D\"{u}sseldorf,
Germany}

\author{H. L{\"o}wen}
\affiliation{Institut f\"ur Theoretische Physik II: Weiche Materie,
Heinrich-Heine-Universit\"at D\"{u}sseldorf, 
Universit{\"a}tsstra{\ss}e 1, D-40225 D\"{u}sseldorf,
Germany}

\date{\today}

%
\begin{abstract}
The influence of hydrodynamic interactions  on lane formation of
oppositely charged driven colloidal suspensions is investigated using
Brownian dynamics computer simulations performed on the Rotne-Prager
level of the mobility tensor.  Two cases are considered, namely
sedimentation and electrophoresis. In the latter case the Oseen
contribution to the mobility tensor is screened due to the opposite
motion of counterions.  The simulation results are compared to that
resulting from simple Brownian dynamics where hydrodynamic
interactions are neglected. For sedimentation, we find that
hydrodynamic interactions strongly disfavor laning. In the
steady-state of lanes, a macroscopic phase separation of lanes is
observed.
This is in marked  contrast to the simple Brownian case where a finite
size of lanes was obtained in the steady-state. For strong Coulomb
interactions between the colloidal particles a lateral square lattice
of oppositely driven lanes is stable similar to the simple Brownian
dynamics.  In an electric field, on the other hand, the behavior is
found in qualitative and quantitative accordance with the case of
neglected hydrodynamics.  
\end{abstract} 
\maketitle
\section{Introduction}
\label{intro}

The dynamics of colloidal particles dispersed in a fluid solvent 
is quite different from the ballistic motion of molecular systems
which is described by Newton's law
\cite{pusey91,loewen_pra_1991,naegele_pr_1996,vermant_jpcm_2005}. 
The viscous solvent both damps 
the motion of a colloidal particle and leads to kicks of the solvent
molecules with the colloidal particle leading to Brownian motion
if the time-scales of the molecular solvent is much faster than that
of the diffusive motion of the colloidal particles. In concentrated suspensions,
a dragged colloidal particle influences the motion of another particles
via the solvent flow field.
These so-called hydrodynamic interaction is typically long-ranged.
While it can be neglected in colloidal suspensions of very small volume fractions,
it induces significant corrections in the equilibrium and nonequilibrium dynamics 
of colloidal suspensions \cite{naegele_pr_1996,pesche_epl_2000}. 
The equilibrium structures and phase boundaries,
on the other hand, are unaffected by hydrodynamic interactions.

In the past years a simple nonequilibrium phase transition
\cite{katz_prb_1983,katz_jsp_1984,schmittmann_1995,loewen_jpcm_2001}
has been discussed in a binary mixture of colloidal particles which
are driven by a constant external field
\cite{dzubiella_pre_2002,netz_epl_2003b,dzubiella_epl_2003,dzubiella_fd_2003}.
The drive is different on the two particle species and could arise
from gravity and from an external electric field in the case of
charged colloidal suspensions.
Brownian dynamics computer simulations with neglected hydrodynamic
interactions strongly support the scenario that - as a function of the
driving strength - the mixtures undergoes a transition from a mixed
steady-state with anisotropic correlations towards a steady-state
where macroscopic lanes are formed.
The transitions has been found for oppositely driven repulsive mixtures
\cite{dzubiella_pre_2002,netz_epl_2003b,dzubiella_epl_2003,dzubiella_fd_2003,delhommelle_pre_2005,pandley_ijmpc_2003,koppl_prl_2006}
in two and three spatial dimensions and it seems to be a first-order
nonequilibrium transition with a significant hysteresis in an order
parameter which detects laning \cite{dzubiella_pre_2002}.
The general scenario occurs also in pedestrian dynamics
\cite{helbing_pa_2006,helbing_njp_2003} and in granular matter
\cite{ehrhardt_pre_2005,coniglio_prl_2005}.

Recently, lane formation was observed in real-space
experiments by Leunissen et al \cite{leunissen_nature_2005}.
Equimolar mixtures of oppositely charged colloidal particles 
were prepared which form binary ionic crystals
\cite{hynninen_prl_2006_b,hynninen_prl_2006_a}.
These crystals were exposed to a strong 
external electric field and the dynamics of lane formation was watched 
by confocal microscopy. Subsequently extensive Brownian dynamics simulations
were carried out  to map the nonequilibrium phase diagram
\cite{rex_pre_2007}. These simulations assumed a Yukawa pair
interaction between the particles
and neglected hydrodynamic interactions completely. 
A wealth of different steady-state structures was detected.
In particular  different orderings were  found 
in the plane perpendicular to the drive including 
a square, triangular or rhombic crystalline lattice of lanes, 
a network structure with a finite structural length and intermediate chain
formation of lanes \cite{rex_pre_2007}.
A rough estimate of the experimental parameters used in Ref.\
\cite{leunissen_nature_2005} reveals that the laned state observed
experimentally indeed falls into the region where laning is expected
to occur by Brownian dynamics computer simulations.

In this paper we address the question how hydrodynamic interactions
influence the scenario and the steady-state diagram of lane formation.
Our motivation to do so is twofold: First, the experiments, of course,
contain hydrodynamic interactions in their full glory, and therefore
an inclusion of hydrodynamic interactions is needed for a quantitative
comparison.
Second, there is a principal need to understand in which direction
hydrodynamics influence lane formation.
In particular it is known \cite{long_epje_2001} that the leading
long-ranged term in the mobility pair tensor is screened if an
electric field is applied since a charged colloidal particles is
surrounded by counterions of opposite charge.
That makes the hydrodynamic interaction significantly different from
e.g. sedimentation induced by different buoyant masses of oppositely
charged colloids where the action of gravity on the microions can
safely be neglected.
In the latter case, the leading part in the mobility tensor at large
interparticle separation is the {\it unscreened} Oseen tensor.
It would be interesting to explore how far the steady-state is
affected by hydrodynamic interactions in both cases of sedimentation
and electrophoresis.

We use Brownian dynamic computer simulations and include hydrodynamic
interaction by using mobility tensors on the Rotne-Prager level.
Both cases of sedimentation and electrophoresis are studied separately
with an unscreened resp.\ screened version of the mobility tensor.
The steady-state phase diagrams and the drift velocity are simulated.
The simulation data are compared to that obtained by simple
Brownian dynamics where hydrodynamic interactions are neglected.
For sedimentation, we find that hydrodynamic interactions strongly
disfavor laning.
In the steady-state of lanes, a macroscopic phase separation of lanes
is observed, i.e.\ the sickness of the lanes are of the system size.
This is in marked contrast to the simple Brownian case where a finite
size of lanes was obtained in the steady-state.
For strong Coulomb interactions between the colloidal particles a
lateral square lattice of oppositely driven lanes is stable similar to
the simple Brownian dynamics.
In an electric field, on the other hand, the behavior is found in
qualitative and quantitative accordance with the case of neglected
hydrodynamics.
All lateral structures are reproduced and the topology of the
steady-state phase diagram is unchanged.

The paper is organized as follows: in section II, we describe our
model and simulation scheme for sedimentation and electrophoresis.
Brownian dynamics simulation results are presented in section III. We
conclude in section IV.

\section{The Model}
\label{sec:model}
We perform Brownian dynamic simulations to study an equimolar binary
mixture of $2N=1024$ oppositely charged colloidal particles of
diameter $\sigma$ dissolved in a solvent fluid of shear viscosity
$\eta$ at temperature $T$ and volume fraction $\phi =
2N\pi\sigma^{3}/6l^{3}$ exposed to an external driving field, where
$l$ is the dimension of a cubic simulation box having periodic
boundary conditions.
Henceforth, $\sigma$ serves as the unit of length and
$k_{\mathrm{B}}T$, the thermal energy, as the energy unit of the
system.
To mimic the experiments by Leunissen et al
\cite{leunissen_nature_2005,royall_jcp_2006} the particles interact
with an effective screened Coulomb potential (or Yukawa potential)
plus a steric repulsion $V_{\mathrm{h}}$:
\begin{equation}
  \label{eq:yukawa_potential}
  V(r_{ij}) = V_{0}\frac{Z_{i}Z_{j}}{(1+\kappa\sigma/2)^2}
  \frac{e^{(-\kappa\sigma(r_{ij}/\sigma-1))}}{r_{ij}/\sigma}+
  V_{\mathrm{h}}(r_{ij})
\end{equation}
with $V_{0}=50k_{\mathrm{B}}T$ the strength of the interaction potential and
$Z_{i}=\pm 1$ the sign of the charge of particle $i$. 
$r_{ij}=|\mathbf{r}_{i}-\mathbf{r}_{j}|$ denotes the distance between
particle $i$ and $j$, where $\mathbf{r}_{i}$ is the coordinate vector.
The inverse screening length $\kappa$ governs the range of the
interaction and is determined by the salt concentration of the
solution.
The steric repulsion between the particles, that prevents the
system from collapsing, is approximated by a repulsive (shifted and
truncated) Lennard-Jones potential
\begin{equation} 
  \label{eq:hardcore_potential}
    V_{\mathrm{h}}(r_{ij}) = \left\{ 
    \begin{array}{ll} 
      \epsilon \left [ \left ( 
      \frac{\sigma}{r_{ij}}\right)^{12} - \left ( 
      \frac{\sigma}{r_{ij}}\right)^{6} +\frac{1}{4}\right] & 
      \textrm{if $r_{ij} \leq 2^{1/6}\sigma$}\\ 
      0 & \textrm{else},
    \end{array} 
    \right. 
\end{equation}
with $\epsilon = 4V_{0}/(1+\kappa\sigma/2)^2$.
The constant external driving field that acts in opposite directions
on the two different particle species reads as
\begin{equation}
  \label{eq:f_ext}
  \mathbf{F}_{i}^{\mathrm{ext}} = Z_{i}f\mathbf{e}_{z},
\end{equation}
where $\mathbf{e}_{z}$ is the unit vector along the $z$ direction and
$f= 150k_\mathrm{B} T/\sigma$ is the strength of the external force.
The external force is supposed to stem from either an electric field
or a gravitational field accompanied with different buoyant masses of
the oppositely charged particles.
Though the external force Eq.\ (\ref{eq:f_ext}) may in both cases be
identical, if the charges and/or buoyant masses are chosen accordingly,
the hydrodynamic interactions are not.

The algorithm used to simulate the diffusive Brownian motion of the
colloidal particles was proposed by Ermak and McCammon
\cite{ermak_jcp_1978}.
Here, the translational displacements of the particles are 
deemed to occur in time steps of fixed length $\Delta t$ and the
update algorithm is given by \cite{allen_tildesley_book}:
\begin{equation} 
  \label{eq:integration_algorithm} 
  \mathbf{r}_{i}(t+\Delta t)=\mathbf{r}_{i}(t)+
  \Delta t \sum_{j=1}^{N}\left\{\frac{\mathbf{D}_{ij}(t)}{k_{\mathrm{B}}T}
  \cdot \mathbf{F}_{j}(t) +
  \mathbf{\nabla}_{\mathbf{r}_{j}}\cdot \mathbf{D}_{ji}(t)\right\} +
  \Delta{\mathbf r}^{{\mathrm G}}_{i}, 
\end{equation}
where $\mathbf{D}_{ij}$ denotes the diffusion tensor field depending on the
positions of the particles at time $t$.
The random displacements $\Delta{\mathbf r}^{{\mathrm G}}_{i}$ are
chosen from a joint Gaussian distribution with mean and covariant
matrix \cite{allen_tildesley_book}
\begin{equation} 
  \label{eq:mean} 
    \langle \Delta{\mathbf r}^{{\mathrm G}}_{i} \rangle_{\mathrm{G}} = 
            0;\;\;\;\; 
    \langle \Delta{\mathbf r}^{{\mathrm G}}_{i} 
            \Delta{\mathbf r}^{{\mathrm G}}_{j} \rangle_{\mathrm{G}} = 
    2{\mathbf D}_{ij}\Delta t,
\end{equation}
where $\langle \ldots \rangle_{\mathrm{G}} $ denotes the average over
the Gaussian noise distribution.
$\mathbf{F}_{i}(t)$, $i=1,...,N$, comprises the nonhydrodynamic
forces due to interparticle interactions, determined by the gradient
of the interaction potentials in Eq.\ (\ref{eq:yukawa_potential}) and
Eq.\ (\ref{eq:hardcore_potential}), and the external force
$\mathbf{F}_{i}^{\mathrm{ext}}$, Eq.\ (\ref{eq:f_ext}), acting onto
particle $i$.

Hydrodynamic interactions are included in the simulation through the
mobility tensor $\bm{\mu}_{ij}=\mathbf{D}_{ij}/k_{\mathrm{B}}T$.
In a first approach we neglect hydrodynamic interactions
completely to asses its effect on the system.
In that case the diffusion tensor is given by Stoke's law
in diagonal form 
\begin{equation}
  \label{eq:without}
  \gamma \bm{\mu}_{ij}=
  \delta_{ij}\mathbf{1},
\end{equation}
with friction $\gamma=3\pi \eta \sigma_{\mathrm{H}}$, where
$\sigma_{\mathrm{H}}$ is the hydrodynamic diameter.
In the sedimentation and electrophoresis situation we approximate the
mobility tensor by two-body interactions.
In this approximation the divergence in Eq.\
(\ref{eq:integration_algorithm}) vanishes always
\cite{wajnryb_pa_2004}.
When studying the sedimentation, the buoyant masses of the oppositely
charged particles are supposed to be such that the same force acts on
the two species but in opposite directions.
The action of gravity on the microions can safely be neglected.
Therefore, in sum no net force is acting on the solvent and overall it
remains quiescent.
Then, for a pair of spheres of hydrodynamic diameter
$\sigma_{\mathrm{H}}$ the mobility tensor is approximated by the
well-known Rotne-Prager expression \cite{rotne_jcp_1969}
\begin{equation} 
  \label{eq:rotne_prager_tensor} 
    \gamma \bm{\mu}^{\mathrm{RP}}_{ij}= 
    \delta_{ij}\mathbf{1}+(1-\delta_{ij})\left[ 
    \frac{3\sigma_{\mathrm{H}}}{8}\mathbb{O}({\mathbf r}_{ij}) + 
    \frac{\sigma_{\mathrm{H}}^{3}}{16}\mathbb{Q}({\mathbf r}_{ij}) 
    \right ],
\end{equation} 
where
\begin{equation} 
  \label{eq:oseen_tensor} 
    \mathbb{O}({\mathbf r}) = \frac{1}{|\mathbf{r}|} 
              ({\mathbf 1}+\hat{\mathbf{r}}\otimes\hat{\mathbf{r}}); \;\;\;\; 
    \mathbb{Q}({\mathbf r}) = \frac{1}{|\mathbf{r}|^{3}} 
              ({\mathbf 1}-3\hat{\mathbf{r}}\otimes\hat{\mathbf{r}}),
\end{equation}
with the unit vector $\hat{\mathbf{r}}=\mathbf{r}/|\mathbf{r}|$,
$\otimes$ a dyadic product, and $\delta_{ij}$ Kronecker's symbol.
On this level of approximation we incorporate all interactions up to
$O((\sigma_{\mathrm{H}}/r)^{3})$.
Higher order contributions such as many-body, coupling between
rotational and translational motions, and lubrication forces are
neglected.
The leading term in Eq.\ (\ref{eq:rotne_prager_tensor}) is given by
$\mathbb{O}({\mathbf r})$ which is of the order of $1/|\mathbf{r}|$
for large distances.

However, when regarding the electrophoresis the mobility tensor
has to be altered since forces induced by the surrounding counterions
into the solvent are in sum equal to the force induced by a colloidal
particle.
Thus, the solvent flow stemming from the drag on the counterions
cannot be neglected as is done in the sedimentation case.
It results in an effective screening of the leading far-distance term
of the hydrodynamic interactions between the colloidal particles as
Long and Ajdai have shown \cite{long_epje_2001}.
Their mobility tensor $\bm{\mu}^{\mathrm{LA}}_{ij}$ reads as
\begin{eqnarray} 
  \label{eq:long_tensor} 
    \nonumber
    \gamma \bm{\mu}^{\mathrm{LA}}_{ij}&=& 
    \delta_{ij}\mathbf{1}+\frac{3\sigma_{\mathrm{H}}}{4} (1-\delta_{ij})
    \left[ 
    \frac{e^{-\kappa r_{ij}}}{r_{ij}} \left(\left(1+\frac{1}{\kappa
    r}+\frac{1}{\kappa^{2}r^{2}}\right) \mathbf{1} -\right.\right.\\
    &&\left.\left.\left(\frac{1}{3}+\frac{1}{\kappa
    r}+\frac{1}{\kappa^{2}r^{2}}\right)
    3\hat{\mathbf{r}}_{ij}\otimes\hat{\mathbf{r}}_{ij}\right)
    + \frac{1}{\kappa^{2}}\mathbb{Q}({\mathbf r}_{ij}) 
    \right ].
\end{eqnarray} 
Here, the leading order term is $\mathbb{Q}({\mathbf r}_{ij})$ which
decays as $1/|\mathbf{r}^{3}|$.
To account for the polymer coating on the colloidal particles that
gives rise to the steric repulsion we choose
$\sigma_{\mathrm{H}}=0.9\sigma$ throughout this paper.

More sophisticated simulation techniques for spherical particles in an
unbounded space including lubrication approximation for particles in
close proximity and multipolar expansion methods are available
\cite{durlofsky_jfm_1987,ladd_jcp_1988,cichocki_jcp_1994,sierou_jfm_2001,banchio_jcp_2003}.
However, in the electrophoresis where the hydrodynamic
interactions of the counterions become important explicit simulations
of all colloidal particles and their counterions -- $110$ per
colloidal particle in the experiments by Leunissen et al -- are still
beyond computational means.
Therefore, we adopted the calculations of Long and Ajdari to our
simulations and compare it to the sedimentation problem on the same
level of accuracy, i.e.\ the Rotne-Prager level.
To our best knowledge this is the first Brownian dynamic simulations
with the Long and Ajdari mobility term \ref{eq:long_tensor}.

Both mobility tensors \ref{eq:rotne_prager_tensor} and
\ref{eq:long_tensor} are long-ranged and thus require an
Ewald-like summation in simulations analogous to Coulomb and
dipole-dipole interactions.
Details on the summation and discussions about appropriate boundary
conditions to the system can be found elsewhere
\cite{smith_pa_1987,smith_fdsc_1987,bradey_jfm_1988b,beenakker_jcp_1986}.
We applied the scheme suggested by Beenakker \cite{beenakker_jcp_1986}
and adapted it for $\bm{\mu}^{\mathrm{LA}}_{ij}$ accordingly.
The square root of the diffusion tensor, needed when calculating the
random displacements in Eq.\ (\ref{eq:mean}), are obtained
from a Cholesky decomposition:
\begin{equation}
  \label{eq:decomposition}
  \mathbf{D} = \mathbf{L}\cdot\mathbf{L}^{\mathrm{T}},
\end{equation}
where $\mathbf{L}$ is a lower triangular matrix and
$\mathbf{L}^{\mathrm{T}}$ is its transpose.
A suitable time scale for our system is $\tau_{\mathrm{B}} =
\gamma \sigma^{2}/k_{\mathrm{B}}T$.
The equations of motion including the external field are
numerically solved using a finite time step $\Delta
t=2\cdot10^{-5}\tau_{\mathrm{B}}$ in all simulations.
Statistics were gathered after an initial relaxation period of
$20\tau_{\mathrm{B}}$.
The starting configuration of all simulations was a homogeneous
mixture.

\section{Results}
\label{sec:results}
\subsection{Order parameter and steady-state phase diagrams}
To assess the effect of hydrodynamic interactions on the
lane behavior of oppositely charged colloidal particles we study a
set of volume fractions $\phi$ and inverse screening lengths
$\kappa^{*}=\kappa\sigma$ and map out nonequilibrium steady-state
phase diagrams for all three situations: hydrodynamic interactions
neglected (A), electrophoresis (B), and sedimentation (C).

A state of lane is thereby identified by a laning order parameter that
is defined through
\begin{equation}
  \label{eq:orderparameter}
    \Phi=\frac{1}{2N}\left
    \langle\sum_{i=1}^{2N}\Phi_{i}\right \rangle_{t},
\end{equation}
where the angular brackets $\langle...\rangle_{t}$ denote a time average. 
The local order parameter $\Phi_{i} =
(n_{\mathrm{l}}-n_{\mathrm{o}})^2/(n_{\mathrm{l}}+n_{\mathrm{o}})^2$
is assigned to every particle $i$, where the numbers $n_{\mathrm{l}}$
and $n_{\mathrm{o}}$ are the number of like charged particles and
oppositely charged particles, respectively, whose projections of
distance onto the plane perpendicular to the field are smaller than a
suitable cut-off length scale $z_{c}$.
$\Phi_{i}$ is equal to $1$ if all particles within this distance
criterion are of the same kind and zero if
$n_{\mathrm{l}}=n_{\mathrm{o}}$, i.e.\ a homogeneous mixture.
We chose for convenience $z_{c}=\frac{3}{4}\sigma$ to detect all lanes
starting from a single queue of particles.
In what follows we will use a threshold: for
$\Phi\geq1/2$ we call the configuration a state of lanes while in the
opposite case ($\Phi<1/2$) we call it a state without lanes.

\begin{figure}[t]
  \begin{center}
  \includegraphics[width=8cm, clip=true, draft=false]{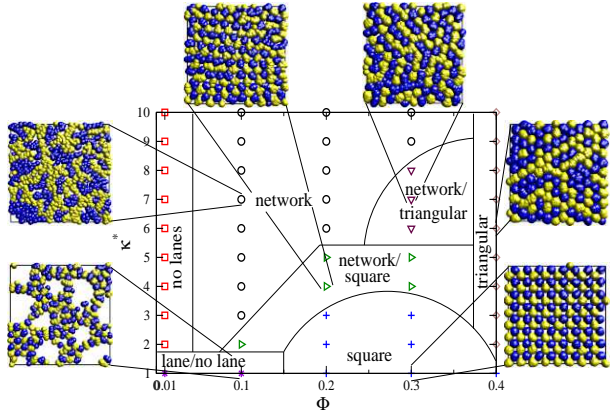}
\caption{(Color online) Nonequilibrium steady-state phase diagram for
  a constant driving force of strength $f= 150k_\mathrm{B} T/\sigma$
  with hydrodynamic interactions neglected accompanied by a typical
  simulation snapshot of the projection of the particle coordinates
  onto the plane perpendicular to the driving field for each different
  state.
  The lines between the phases are a guide for the eye.}
\label{fig:phasediagram1}
  \end{center}
\end{figure}

\begin{figure}[t]
  \begin{center}
  \includegraphics[width=8cm, clip=true, draft=false]{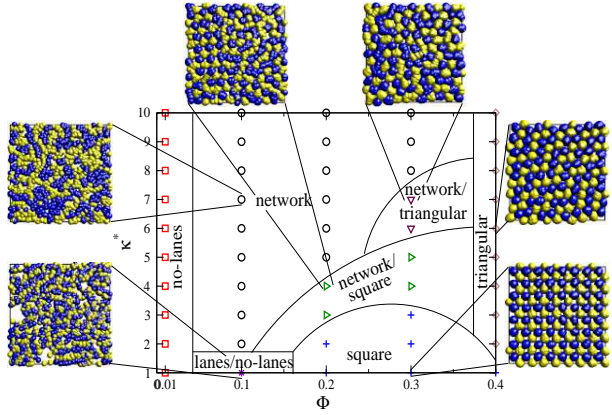}
\caption{(Color online) Same as Fig. \ref{fig:phasediagram1}
  but for electrophoresis with hydrodynamic interactions taken into
  account through $\bm{\mu}_{ij}^{\mathrm{LA}}$ in Eq.\
  (\ref{eq:long_tensor}).
  The phase diagram reveals only minor differences as compared to the
  the case of neglected hydrodynamic interactions in Fig.\
  \ref{fig:phasediagram1}.}
\label{fig:phasediagram2}
  \end{center}
\end{figure}

\begin{figure}[t]
  \begin{center}
  \includegraphics[width=8cm, clip=true, draft=false]{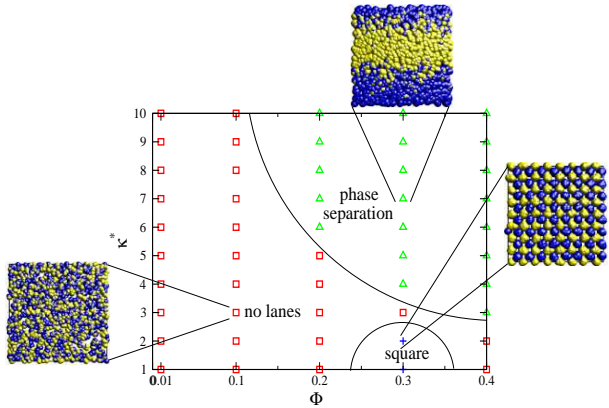}
\caption{(Color online) Same as Fig. \ref{fig:phasediagram1}
  but for sedimentation with hydrodynamic interactions taken into
  account through $\bm{\mu}_{ij}^{\mathrm{RP}}$ in Eq.\
  (\ref{eq:rotne_prager_tensor}).
  The phase diagram shows significant changes as compared to Fig.\
  \ref{fig:phasediagram1} and Fig.\ \ref{fig:phasediagram2}.}
\label{fig:phasediagram3}
  \end{center}
\end{figure}

We observe that lanes form different structures in the plane
perpendicular to the driving direction for different values of
$\kappa^{*}$ and $\phi$.
We find lanes placed on a square or triangular lattice, a network-like
structure (reminiscent of a  bicontinous microemulsion or
microphase-separated system), coexistence regimes of the same, and
macroscopically separated lanes.
The resulting nonequilibrium steady-state phase diagrams are shown in
Figures \ref{fig:phasediagram1}, \ref{fig:phasediagram2}, and
\ref{fig:phasediagram3}.
They are accompanied with typical simulation snapshots of the
projection of all particle coordinates onto the $xy$-plane of the
respective situation.

What can be seen at first sight is that the qualitative behavior of
situation (A) and (B) in Figures \ref{fig:phasediagram1} and
\ref{fig:phasediagram2} is almost identical with only subtle
differences, while on the other hand the phase behavior changes
drastically for situation (C), Figure \ref{fig:phasediagram3}.
In the latter the whole phase diagram is altered and the diversity of
phases found is reduced compared to the first two cases.

\subsection{Comparison of simulation results for neglected
hydrodynamic interactions and electrophoresis}
\label{sec:el}
In this subsection we briefly describe the two phase diagrams in
Figures \ref{fig:phasediagram1} and \ref{fig:phasediagram2} and their
differences, beginning at low volume fractions and ending at high
ones, and then dwell on the third diagram, Fig.\
\ref{fig:phasediagram3}, thereafter in subsection \ref{sec:sed}.
A more ample discussion on how different phases are identified and
what structural correlations they exhibit can be found in a previous
work of the authors on the same system with hydrodynamic interactions
neglected but for a slightly different driving strength and larger
systems \cite{rex_pre_2007}.
Here, we find for situation (A) virtually the same results as in the
previous work with only one difference, namely that we do not
encounter a rhombic phase for $\phi=0.4$ and $\kappa^{*}=1,2,3$.

For very low volume fraction, $\phi\lesssim 0.01$, in both systems the
correlations between the particles are not sufficient to form lanes at
all.
Thus, the systems are in a phase of no-lanes.
Only for very low salt concentration, i.e.\ small $\kappa^{*}$, where
the electrostatic coupling between the colloidal particles is strong,
we find a coexistence region between lanes and no-lanes.
Here, the region with no-lanes consists of voids, where hardly any
particle is found.
The structure of the lane region, on the other hand, is different in the
two situations.
For situation (A) the corresponding snapshot in Fig.\
\ref{fig:phasediagram1} reveals fixed lattice points while the
snapshot in Fig.\ \ref{fig:phasediagram2} situation (B) shows a
network-like structure.
For situation (A), an initial configuration with lanes placed on a square
lattice separated from a completely depleted region is stable in
simulations, as well.
Thus, we assume that in situation (A) the lanes/no-lanes phase is a
transient state toward a complete square lattice and no-lane phase
separation.
Hydrodynamic interactions destroy the coexistence phase for $\phi=0.01$.
It only occurs in a denser system with $\phi=0.1$ whereas in situation
(A) this state shows already up at $\phi=0.01$.
Additionally, the voids are more pronounced in the latter case
compared to situation (B).
Upon increasing $\kappa^{*}\geq2$ for $\phi=0.1$ in both situations we
find a network-like structure whose characteristic spacing is
increasing with increasing $\kappa^{*}$.
For situation (A) there is also a small coexistence region between
network and square lattice at $\kappa^{*}=2$.

To obtain a quantitative measure of the characteristic spacing in the
network structure we determine a structure factor perpendicular to the
driving field of like-charged particles.
The steady-state partial structure factor has been calculated by
evaluating the expression 
\begin{equation}
  \label{eq:Sofk}
    S_{\mathrm{AA}}(k) = 1+\rho\hat{h}_{\mathrm{AA}}(k),
\end{equation}
with $\rho=6\phi/\pi\sigma^{3}$ the number density and the wave vector
$k=|\mathbf{k}|$, where $\mathbf{k}=(2\pi/l)(k_{x},k_{y})$ and
$k_{x}$, $k_{y}$ are integers.
$\hat{h}_{\mathrm{AA}}(k)$ is the Fourier transform of the total correlation
function $h_{\mathrm{AA}}(\mathbf{r}_{\perp})=g_{\mathrm{AA}}(r_{\perp})-1$ with
$\mathbf{r}=(\mathbf{r}_{\perp},z)$ and
\begin{equation}
  \label{eq:gofrperp}
    g_{\mathrm{AA}}(r_{\perp}) = \frac{1}{\rho N}
      \left \langle
      \sum_{\substack{i,j\\ (Z_{i}=Z_{j},\; i\ne j)}}^{2N}
      \delta(\mathbf{r}_{\perp}-|\mathbf{r}_{i\;\perp}-\mathbf{r}_{j\;\perp}|)
      \delta(z_{i}-z_{j})
    \right\rangle_{t},
\end{equation}
where $\delta(\mathbf{x})$ denotes Dirac's delta-distribution.
\begin{figure}[t]
  \begin{center}
  \includegraphics[width=8cm, clip=true, draft=false]{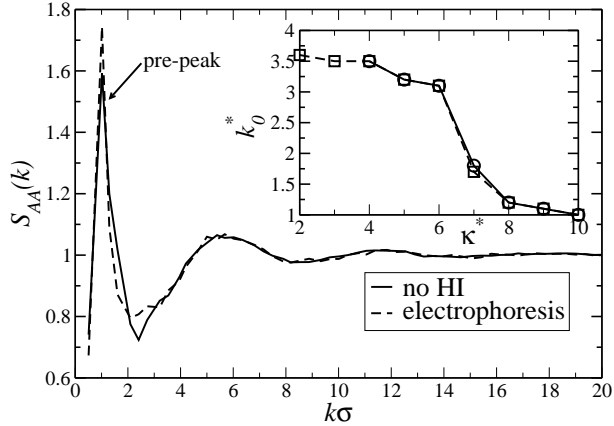}
\caption{Partial structure factor $S_{\perp}(k)$ of like-charged
particles perpendicular to the driving field for $\kappa^{*}=10$ and
$\phi=0.1$ for hydrodynamics neglected and electrophoresis.
The pre-peak at $k^{*}_{0}=k_{0}\sigma$ indicates an additional length
scale of the structure in the network-like phase.
The inset shows the position $k^{*}_{0}$ of the pre-peak as a function
of the inverse screening length $\kappa^{*}$ for a fixed volume
fraction of $\phi=0.1$.}
\label{fig:sk}
  \end{center}
\end{figure}
An example of the steady-state partial structure factors for
$\kappa^{*}=10$ and $\phi=0.1$ for both situation, (A) and (B), is
shown in Fig. \ref{fig:sk}.
One clearly observes a pronounced pre-peak at the wave number
$k_{0}$ in both cases.
A pre-peak in the structure factor is an indication of an additional
mesoscopic length scale as typical for bicontinous networks, such as
e.g.\ microemulsions \cite{teubner_jcp_1987,gompper_prl_1989}.
In the inset we additionally present the position $k_{0}$ of the
pre-peak as a function of the inverse screening length.
It is evident from the picture that the characteristic spacing is
indeed growing with increasing $\kappa^{*}$.
We find hardly any difference between situation (A) and (B).

For $\phi=0.2$ an additional phase for small inverse screening length
shows up in both phase diagrams.
Oppositely driven lanes are placed on a square lattice with an
alternating charge pattern.
The formation of this lattice structure can be qualitatively
understood from an effective interaction between oppositely charged
driven lanes which has a short-ranged repulsive and a long-ranged
attractive interaction.
The former stems from the friction between oppositely driven
particles while the later results form the Coulomb interaction.
The square lattice then reduces the electrostatic energy of the system 
because each particle has only oppositely charged neighbors.
For increasing salt concentrations we encounter a coexistence region
between the square lattice and the network-like phase and finally end up
in a pure network-like phase.
The phase diagram is in both situations very similar, only the borders
of the transitions are slightly shifted.
In the electrophoresis case the network-like structure is preferred to
the square lattice.

For a higher volume fractions of $\phi=0.3$ a coexistence regime
between a triangular lattice and a network-like structure is found.
The lattice-points in the triangular phase are rather randomly
decorated with different charges.
Here, the short range repulsion plays the dominant role compared to
the electrostatic interaction.
It enforces a triangular lattice due to packing effects although
electrostatically it is strongly disfavored because like-charged
particles necessarily occupy lattice points next to each other.
Again, hydrodynamic interactions slightly shift the phase boundaries
towards the network-like structure.

For the highest volume fraction studied, $\phi=0.4$, both phase
diagrams show exactly the same behavior.
Here, the short range repulsions dictates the phase behavior for
nearly all salt concentration but for $\kappa^{*}=1$ and enforces
lanes to be placed on a triangular lattice.
Only for $\kappa^{*}=1$, where electrostatic interactions are
prominent, a square lattice is preferred.
In principle a square lattice is possible up to the packing of a
simple cubic lattice of $\phi=0.52$.

In summary we observe very similar behavior in both situations.
The observed differences can be qualitatively explained by the fact
that hydrodynamic interactions disfavor lanes driven oppositely past
each other.

\subsection{Sedimentation}
\label{sec:sed}
Regarding sedimentation, Fig.\ \ref{fig:phasediagram3}, the whole
phase diagram exhibits {\it only} three different phases.
For volume fractions $\phi \leq 0.1$ we do not find lane formation for
all inverse screening length studied.
For increasing volume fractions and strong electrostatic interactions,
$\kappa^{*}\leq2$, first the square lattice at $\phi\approx0.3$, that
is also present in the previous two situations, is recovered and then
the system reenters a region with no-lanes for $\phi=0.4$.
This behavior nicely illustrates the competition between hydrodynamic
interactions disfavoring lanes driven oppositely past each other and
the electrostatic interactions favoring a square lattice.
Only for the small regime around $\phi\approx0.3$ the electrostatics
succeeds the hydrodynamic interactions and enforces a square lattice.
For all other volume fractions laning is destroyed.
However, for stronger salt concentrations, where the Coulombic
coupling is reduced, we discover a situation that
is not present in the previous situations (A) and (B), namely a region
in which only two big completely separated lanes.
We call this state {\it phase separated}.
In that case the long ranged hydrodynamic interactions prescribe the
structure and the short ranged Yukawa interaction plays its role only
at the rough interface of the two phases.
From our simulations we conclude the lanes are separated by half of the
box length.

\subsection{Drift velocity}
\label{sec:vd}
Now, we study the influence of hydrodynamic interaction on the drift
velocity along the field direction that is defined as follows
\begin{equation}
  \label{eq:vel}
  v^{2}:=\lim_{t\rightarrow\infty}\frac{\left\langle\left[(\mathbf{r}_{i}(t)
  - \mathbf{r}_{i}(0))\cdot\mathbf{e}_{z}\right]^{2}\right\rangle}
  {t^{2}}.
\end{equation}
This entity measures the mean-square displacement of each particle
in the nonequilibrium steady-state.
A study on the effect of hydrodynamics on the drift velocity of like
charged colloidal particles was carried out by Watzlawek and N\"agele
\cite{watzlawek_jcis_1999}.
We study two cases, first we fix the volume fraction at $\phi=0.3$ and
vary the inverse screening length and afterward vice versa for
$\kappa^{*}=1$.

\begin{figure}[t]
  \begin{center}
  \includegraphics[width=8cm, clip=true, draft=false]{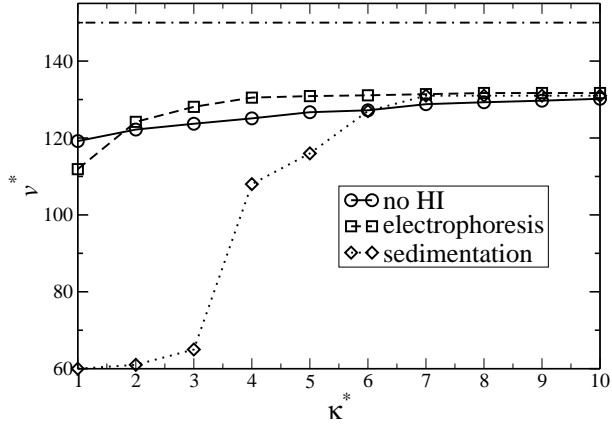}
\caption{Average dimensionless drift velocity
  $v^{*}=v\tau_{\mathrm{B}}/\sigma$ in drive direction as a function
  of the inverse screening length $\kappa^{*}$ at $\phi=0.3$ for
  Brownian dynamic simulations with hydrodynamic interactions
  neglected, taken into account through $\bm{\mu}_{ij}^{\mathrm{LA}}$,
  and $\bm{\mu}_{ij}^{\mathrm{RP}}$.}
\label{fig:v}
  \end{center}
\end{figure}

In Fig.\ \ref{fig:v} we display $v^{*}=v\tau_{\mathrm{B}}/\sigma$ for
a fixed volume fraction $\phi=0.3$ as a function of the inverse
screening length $\kappa^{*}$ for all three situations.
For all cases the drift velocity increases with
decreasing Coulomb coupling because oppositely charged colloids
attract each other while driven in opposite direction and lanes
mutually retard each other.
For very strongly screened particles where this friction is less
important all three curves reveal approximately the same value of
$v\approx 130\tau_{\mathrm{B}}/\sigma$.
Accordingly, this value is close to the drift velocity of $v =
150\tau_{\mathrm{B}}/\sigma$ for a system of infinite dilution
subjected to the same driving force.
While in (A) and (B) $v$ grows gradually, in the sedimentation curve
we encounter a jump in the drift velocity between $\kappa^{*}=3$ and
$\kappa^{*}=4$.
This coincides with the transition from the no-lane regime to the phase
separated regime, see the phase diagram Fig.\ \ref{fig:phasediagram3}.
On the other hand for $\kappa^{*}\leq2$, where we find a square lattice,
the drift velocity is similar to $\kappa^{*}=3$.
From that we conclude that the phase separated state of lanes supports
particle transport while lanes placed on a square lattice enforced by
strong Coulombic interactions slows down particle transportation.
A further interesting feature is that curves for (A) and (B)
intersect between $\kappa^{*}=1$ and $\kappa^{*}=2$ and that the screened
hydrodynamic interaction enhance the drift velocity for larger inverse
screening length.
The same is true for the unscreened hydrodynamic interactions in the
sedimentation for $\kappa^{*}\geq6$.
\begin{figure}[t]
  \begin{center}
  \includegraphics[width=8cm, clip=true, draft=false]{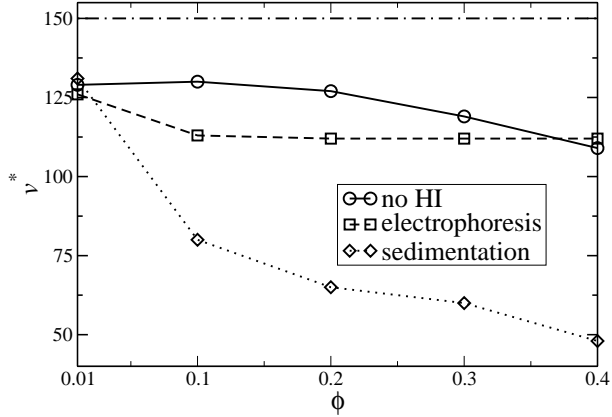}
\caption{Average dimensionless drift velocity
  $v^{*}=v\tau_{\mathrm{B}}/\sigma$ in drive direction as a function
  of the volume fractions $\phi$ at $\kappa^{*}=1$ for
  Brownian dynamic simulations with hydrodynamic interactions
  neglected, taken into account through $\bm{\mu}_{ij}^{\mathrm{LA}}$,
  and $\bm{\mu}_{ij}^{\mathrm{RP}}$.}
\label{fig:v2}
  \end{center}
\end{figure}
When studying the drift velocity for a fixed inverse screening length
but for varying volume fractions in Fig.\ \ref{fig:v2} we find again
an intersection point of the curves for situations (A) and (B).
Here, the drift velocity in the electrophoresis reaches an
approximately constant value of $v^{*}\approx112$ for $\phi=0.1 - 0.4$
whereas it gradually decreases when hydrodynamic interactions
are neglected.
For the sedimentation the drift velocity decreases monotonically.
In contrast to the case of varying salt concentration we do not
encounter a jump in the sedimentation drift velocity when entering the
square lattice at $\phi=0.3$ and reentering the no-lane regime at
$\phi=0.4$.

\section{Conclusions}
In conclusion the influence of hydrodynamic interactions on lane
formation of opposite charged colloids driven by an electric field or
by gravity was investigated by Brownian dynamics computer
simulations.
Hydrodynamic interactions were included on the Rotne-Prager level.
For an electric field, the leading Oseen term is screened due to the
presence of counterions.
The latter fact has lead to very similar steady state phase diagrams
for an electric field as a driving source than that in the simple case
of neglected hydrodynamic interactions.
Various steady state were obtained as a function of the colloidal
density and the range of the interaction.
They can qualitatively be understood in terms of a competition of the
mutual Coulomb attraction and friction of sliding lanes.
At high densities the lateral structure is crystalline, the crystal is
either triangular as dictated by packing at high densities and high
screening or square-like at low-screening which minimizes the Coulomb
attractive energy.
On the other hand, in sedimentation where the two colloidal species
have the same buoyant mass up to a relative sign, friction of sliding
lanes is strongly enhanced leading to macroscopic separation of
lanes.

The steady-state phase diagram can in principle be verified in
real-space experiments of charged suspensions which are driven in an
electric field or sedimenting \cite{leunissen_nature_2005}.
It would be interesting to construct a microscopic theory for the lane
transitions which includes the lateral crystalline structure.
The instability analysis within a dynamical density functional theory
as applied to the case of equal charges in two spatial dimensions
\cite{dzubiella_epl_2003,dzubiella_pre_2004} should in principle be
generalizable to the case of oppositely charged particles.

Finally more sophisticated simulations schemes are needed in order to
go beyond the Rotne-Prager level of approximation used in this paper.
Among the various promising approaches are the stochastic rotation
dynamics code \cite{padding_pre_2006,padding_prl_2004}, a lattice
Boltzmann theory including hydrodynamics
\cite{lobaskin_njp_2004,chatterji_jcp_2005,capuani_jcp_2006,chatterji_jcp_2007,lobaskin_prl_2007}
and counterion flow or the recently developed fluid particle dynamics
methods \cite{kodama_jpcm_2004,kim_prl_2006,kim_mts_2005}.

\section*{ACKNOWLEDGMENTS}

We thank M. Leunissen, A. van Blaaderen, R. Yamamoto, A. Louis, and J.
Padding for helpful remarks and the DFG (SFB TR6, project section D1)
and the Graduiertenf{\"o}rderung of the Heinrich-Heine-Universit\"at
D\"usseldorf for financial support.


\end{document}